\documentclass[english,aps,prl,twocolumn,showpacs,superscriptaddress]{revtex4}

\usepackage{bm}
\usepackage{appendix}
\usepackage{graphicx}
\usepackage{amsmath}
\usepackage{amssymb}
\usepackage{color}
\usepackage{lipsum}
\usepackage[colorlinks=true,citecolor=blue,urlcolor=blue]{hyperref}
\begin{document}
\title{Chiral Dynamics Near Intra- and Inter-Band Exceptional Points under Dissipative Spin-Orbital-Angular-Momentum Coupling}
\date{\today}

\author{Bo-Wen Liu}
\author{Ke-Ji Chen}
\email{chenkeji2010@gmail.com}
\affiliation{Key Laboratory of Quantum States of Matter and Optical Field Manipulation of Zhejiang Province, Department of Physics, Zhejiang Sci-Tech University, 310018 Hangzhou, China}

\author{Fan Wu}
\email{t21060@fzu.edu.cn}
\affiliation{Fujian Key Laboratory of Quantum Information and Quantum Optics, College of Physics and Information Engineering, Fuzhou University, Fuzhou, Fujian 350108, China}

\author{Wei Yi}
\email{wyiz@ustc.edu.cn}
\affiliation{Laboratory of Quantum Information, University of Science and Technology of China, Hefei 230026, China}
\affiliation{Anhui Province Key Laboratory of Quantum Network, University of Science and Technology of China, Hefei, 230026, China}
\affiliation{CAS Center For Excellence in Quantum Information and Quantum Physics, Hefei 230026, China}
\affiliation{Hefei National Laboratory, University of Science and Technology of China, Hefei 230088, China}

\begin{abstract}
We study the parametric chiral dynamics of atoms under dissipative spin-orbital-angular-momentum coupling (SOAMC).
With atoms confined in the ring-shaped potential of the Laguerre-Gaussian Raman beams, the SOAMC not only couples the atomic center-of-mass angular momentum to the hyperfine spins, but also mixes different bands in the radial direction. 
This gives rise to a series of exceptional points of two types, the intra-band and the inter-band. 
Leveraging the topology of the spectral Riemann surface close to these exceptional points, we demonstrate the path-dependent chiral transfer of atoms to the higher-lying bands, by evolving the system along closed loops in the parameter space.
Specifically, we illustrate two distinct scenarios,  
characterized by different mechanisms, where the atoms can be transferred to designated SOAMC-dressed bands.
Our work demonstrates the rich exceptional structure in atom gases under dissipative SOAMC, and offers a novel route toward populating higher bands.
\end{abstract}
\maketitle

Synthetic spin-orbital-angular-momentum coupling (SOAMC) represents a latest addition to the toolbox of quantum simulation in cold atoms~\cite{ Lin-prl-18,Jiang-prl-19,Peng-aapps-22}. Like the synthetic spin-orbit coupling before it~\cite{socreview1,socreview2,socreview3,socreview4,socreview5,socreview6}, SOAMC couples the atomic center-of-mass motion to their internal hyperfine-spin degrees of freedom, and, by modifying the single-particle dispersion~\cite{Pu-pra-15, Qu-pra-15, Sun-pra-15, Chen-pra-16}, enables the simulation of topological matter~\cite{Pu-prl-20,Chen-prr-22}, interaction-induced phase transitions~\cite{Hu-prr-19,Chen-pra-19,Duan-pra-20,li-pra-22,Prikhodko-pra-22}, vortices in superfluids~\cite{Chen-20,Wang-21}, as well as exotic few- and many-body pairing states~\cite{Han-22, Chen-24, Peng-aapps-22}.

Alternatively, the single-particle dispersion can also be dramatically modified by dissipation. 
For instance, the eigenspectrum of a system can be made complex by introducing non-Hermiticity, often through post selection in the open-system dynamics. A whole new range of non-Hermitian features, such as the exceptional points (EPs)~\cite{Bender-07, EP-review1}, the non-Hermiticity enriched symmetry~\cite{Ueda-review1,Ueda-review2, Bergholtz-21}, and non-Hermitian skin effects~\cite{Wang-prl-18, Kunst-prl-18} then give rise to complex spectral geometries and intriguing dynamics that have stimulated tremendous activities of late~\cite{Rotter-16,Chan-18, Yang-19, Liu-21, Song-21,Murch-21,Murch-22,Wei-22, Jo-22,Nasari-22,Yan-prl-22,Rotter-pra-15, Hassan-prl-17,Sun-pra-23,Qi-prl-24}. 
Outstanding examples include the chiral state transfer near EPs in lossy cold atoms~\cite{Jo-22,Sun-pra-23}, and 
topological edge states associated with the non-Bloch topology~\cite{Yan-prl-22}.
Nevertheless, while the quantized angular momenta and multiple radial modes typical of SOAM-coupled gases have much potential in quantum control and engineering, the spectral geometry and its dynamic consequences under a non-Hermitian SOAMC have yet to be explored.

In this work, we study the parametric chiral dynamics of cold atoms under a non-Hermitian SOAMC.
The SOAMC couples both the quantized angular momenta and radial modes (or bands) of atoms in a ring-shaped potential, leading to a complicated spectral Riemann surface with a series of intra- and inter-band EPs in the parameter space.
We show that these EPs are useful for the chiral population of higher band. Specifically, by varying system parameters along a closed loop near the EPs (dubbed EP encircling), the atoms can end up in a designated high-lying band, provided the loop is traversed with the required chirality. 
Based on the geometry of the Riemann surface, we demonstrate two distinct scenarios, wherein such chiral state transfer can be achieved either non-adiabatically or adiabatically, and involving multiple EPs.
Our study is experimentally accessible, and has further implications for quantum control and simulation with cold atoms.

\begin{figure}[t]
\begin{center}
\includegraphics[width=0.44\textwidth]{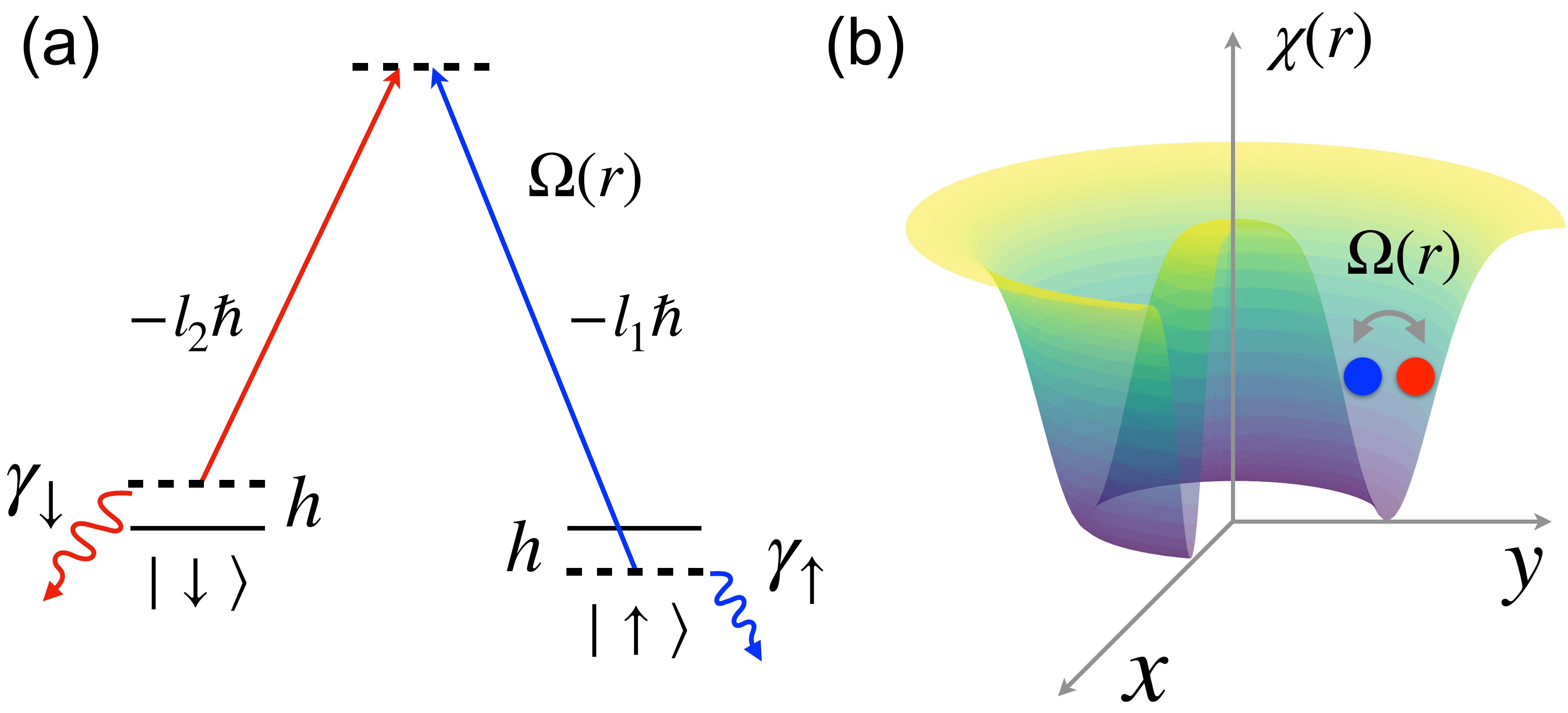}
\caption{(a) Schematic level structure with spin-dependent atom loss. (b) The ac Stark shift of the Raman process $\chi (r)$ provides the ring-shaped confinement, while the Raman coupling $\Omega(r)$ drives transitions between distinct hyperfine states.}
\label{Fig1}
\end{center}
\end{figure}

{\it \color{blue}{Model.---}}
We consider an ensemble of non-interacting two-component atoms confined in the two-dimensional $x-y$ plane. The SOAMC is generated by a two-photon Raman process using a pair of co-propagating Laguerre-Gaussian beams carrying angular momenta $-l_1 \hbar$ and $-l_2 \hbar$, as illustrated in Fig.~\ref{Fig1}(a). The resulting  Raman field features an inhomogeneous Raman coupling $\Omega(r)$ and a phase winding $e^{-2il \theta}$ with $2l=l_1-l_2$ in the polar coordinates ${\bf r}=(r,\theta)$.  Following Ref.~\cite{Jo-22}, we consider laser-assisted spin-dependent loss, with rates $\gamma_{\uparrow, \downarrow}$.  
The effective non-Hermitian Hamiltonian becomes (under a gauge transformation)~\cite{Jiang-prl-19}:  $H_0=\int{ d \bf r} \psi^{\dag}{\cal H}_s \psi$,  where  $\psi=\left(\psi_{\uparrow}, \psi_{\downarrow}\right)^{T}$ denotes the two-component atomic field and ${\cal H}_s$ is (up to a constant term)
\begin{align}
{\cal H}_s =&-\frac{\hbar^2}{2M}\frac{1}{r}\frac{\partial}{\partial r}\left(r\frac{\partial}{\partial r}\right)+\frac{\left(L_z-l \hbar \sigma_z\right)^2}{2Mr^2} \nonumber\\
&+\chi(r)+\Omega (r)\sigma_{x}-\omega \sigma_z,
\label{Hs}
\end{align}
with  $M$ the atom mass and $\sigma_{j}(j=x,y,z)$ the Pauli matrices.  Here $\omega=h-i \gamma$, with $h$ the two-photon detuning and $\gamma=(\gamma_{\downarrow}-\gamma_{\uparrow})/2$.  Under the Raman-assisted SOAMC, the hyperfine spin states are coupled to the atomic orbital angular momentum through the term
 $-l \hbar L_{z}\sigma_{z}/(Mr^2)$ with $L_z=-i \hbar \partial /\partial \theta$. The ac Stark shift $\chi (r)$ of the Raman beams provides
a ring-shaped trapping potential~[see Fig.~\ref{Fig1}(b)].  The  Raman coupling and the ac Stark shift take the forms $\Omega(r)=\Omega_0 I(r)$ and $\chi(r)=\chi_0 I(r)$, respectively, with $\Omega_0$ and $\chi_0$ their corresponding strengths. The intensity profile is $I(r) =   (\sqrt{2}r/w)^{|l_1|+|l_2|}e^{-2 r^2 /w^2}$, and $w$ the beam waist. 
For numerical calculations below, an isotropic hard-wall boundary with radius $R$ is introduced.  

\begin{figure}[t]
\includegraphics[width=0.46\textwidth]{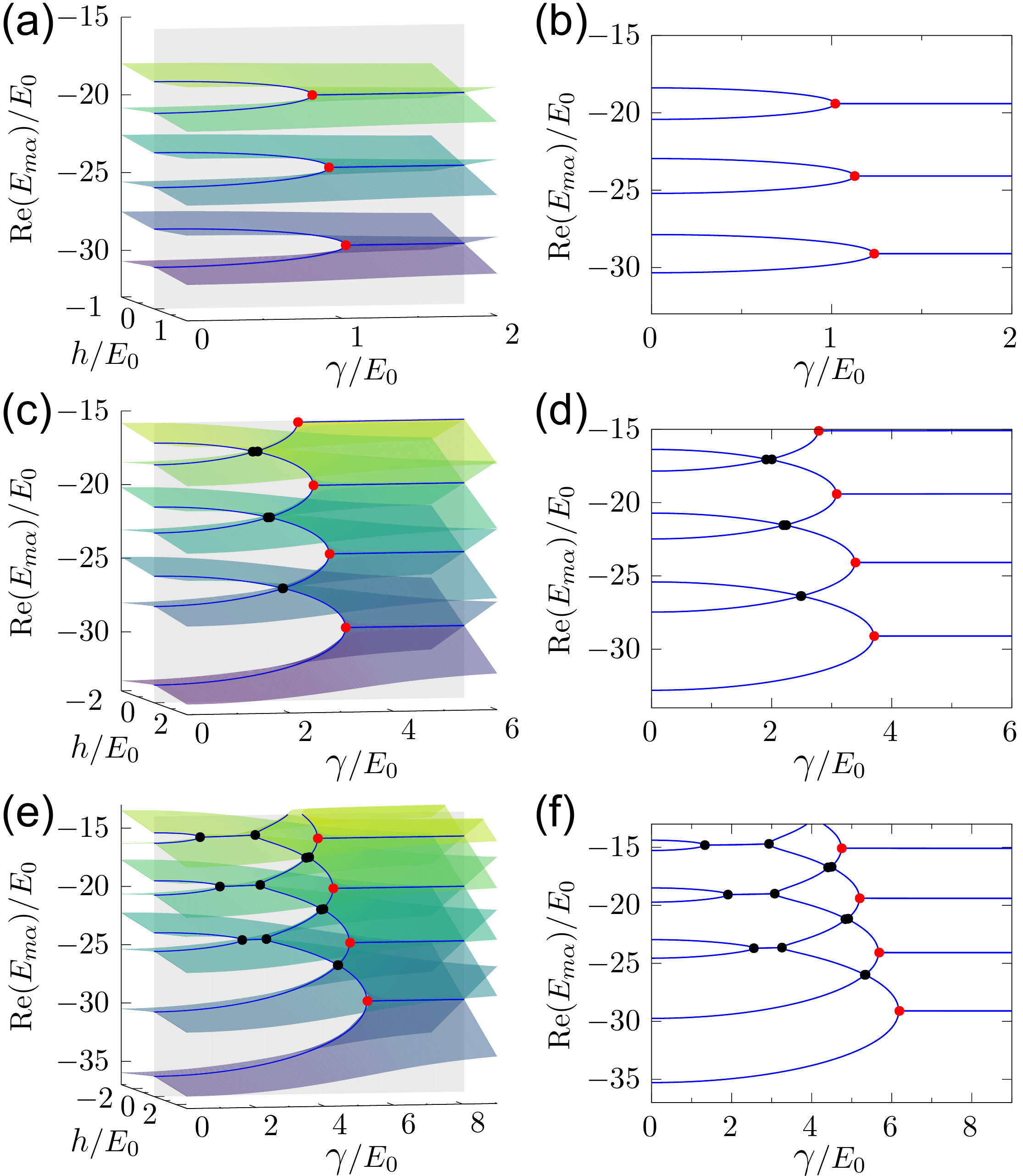}
\caption{Riemann surfaces of the real components of the eigenspectra,  parameterized by $h$ and $\gamma$ in the $m=0$ sector.  The right panels show the cross sections of the Riemann surfaces on
the plane of  $h=0$. The blue solid lines indicate the branch cuts, and red (black) dots mark intra-(inter-) band Eps. (a)(b) $\Omega_{0}/E_{0}=0.27$, (c)(d) $\Omega_{0}/E_{0}=0.82$, and (e)(f) $\Omega_{0}/E_{0}=1.37$.  Other parameters are fixed as $l_{1}=4$, $l_{2}=-4$, $w=3a$, $R=15a$, and $\chi_{0}/E_{0}=-6.85$. We take $E_{0}=\hbar^{2}/(Ma^{2})$  as the unit of energy.}
\label{Fig2}
\end{figure}

The eigenspectrum of Hamiltonian~(\ref{Hs}) can be numerically obtained under an appropriate choice of basis~[see Supplemental Material]. Here, 
we adopt an alternative approach to clarify the underlying physics. Since $[L_z, {\cal H}_s]=0$, we expand the field operator as $ \psi_{\sigma}({\bf r})=\sum_{m,n} \varphi_{mn \sigma}(r) \Theta_m(\theta) a_{mn \sigma} $, with $\Theta_{m}(\theta)=e^{i m \theta}/\sqrt{2\pi}$ and $a_{mn \sigma}$  the corresponding annihilation operator.  The radial function  satisfies 
$K_{\sigma}({\bf r}) \varphi_{m n \sigma}(r)\Theta_m(\theta)=\epsilon_{mn\sigma}\varphi_{mn\sigma}(r)\Theta_m(\theta)$, where $K_{\sigma}({\bf r})=-\frac{\hbar^2}{2M}\Big[\frac{1}{r}\frac{\partial}{\partial r}\left(r\frac{\partial}{\partial r}\right)+\frac{1}{r^2}\left(\frac{\partial}{\partial \theta}-i l \tau \right)^2\Big]+\chi_0 I(r)$, and  $\tau=+1(-1) $ for  $\sigma=\uparrow(\downarrow)$. Here  $m \in {\mathbb Z}$ and $n\in {\mathbb N}$ denote the angular and radial quantum numbers, respectively. For a sufficiently deep ac Stark potential~[see Fig.~\ref{Fig1}(b)], adjacent radial levels for a given angular quantum number $m$ are well separated,  we therefore identify  $n$ as an effective band index ($n=0,1,2,3, \cdots$ corresponding to $s,p,d,f, \cdots$).   The Hamiltonian can then  be written as $H_0=\sum_{m}H_{m}$, with 
\begin{align}
\label{Hamiltonian-eff}
H_{m}=&\sum_{n\sigma } \Big[\epsilon_{mn\sigma}-\omega \tau \Big]a^{\dag}_{mn \sigma}a_{mn\sigma}\nonumber\\
&+\sum_{nn'}\Big[\Omega^{m}_{n\uparrow, n' \downarrow} a^{\dag}_{mn \uparrow}a_{m n' \downarrow}+h.c.\Big],
\end{align}
where $\Omega^{m}_{n \sigma, n' \sigma'} =  \Omega_0 \int rdr \varphi_{m n \sigma} I(r) \varphi_{m n' \sigma '}$ couples spin components in different radial bands. Diagonalizing  $H_{m}$ yields the eigenvalues $E_{m\alpha}$, where $
\alpha \in \mathbb{Z}^{+}$ labels the SOAMC-dressed bands, ordered in accordance with the band indices in the presence of SOAMC.
Specifically, $\alpha=2n+1$ ($\alpha=2n+2$) denotes the lower (upper) branch of the $n$th dressed band. We show in Fig.~\ref{Fig2} typical real
components of the eigenspectra in the $m=0$ sector;   whereas those in other $m$ sectors are demonstrated in the
Supplemental Material.  As illustrated in  Fig.~\ref{Fig2}(a)(b),  under a weak Raman coupling, different dressed bands are indeed well separated, but each features an intra-band EP (marked in red).  With increasing Raman coupling strength, different dressed bands hybridize, leading to the emergence of inter-band EPs [see Fig.~\ref{Fig2}(c)(d), marked in black; see Supplemental Material for the imaginary bands]. 
The number of EPs increases with the coupling strength [see Fig.~\ref{Fig2}(e)(f)], and the coexistence of intra- and inter-band EPs gives rise to a complicated spectral
Riemann surface, which have rich dynamic implications.

\begin{figure}[t]
\begin{center}
\includegraphics[width=0.46\textwidth]{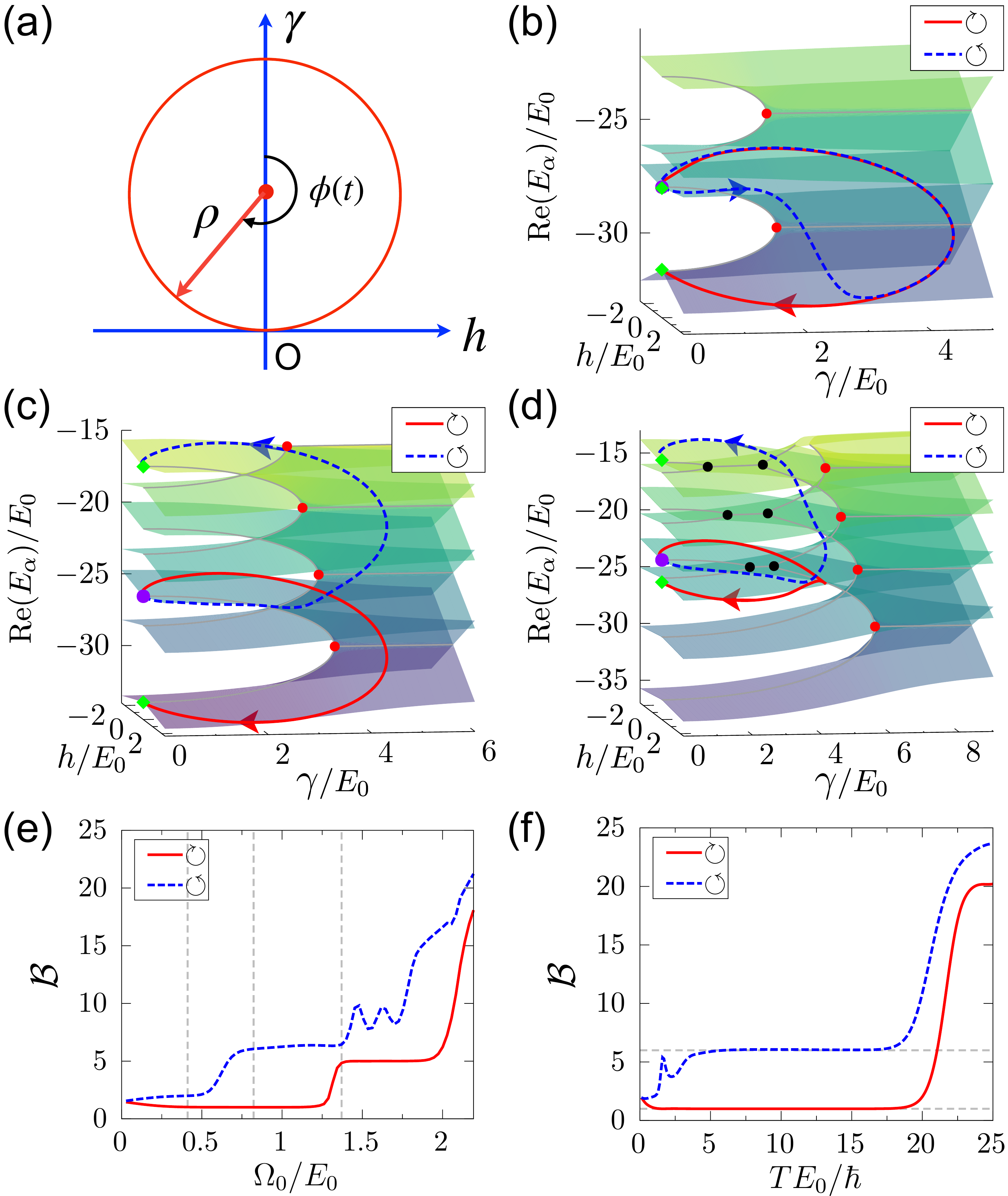}
\caption{(a) Schemaic evolution of $h$ and $\gamma$. Here, $\phi(t)=\pm 2\pi t /T+\pi$, where $+$ ($-$) corresponds to clockwise  (counterclockwise) encircling.  (b)(c)(d) The spectral trajectories under clockwise ($\circlearrowright$, red solid) and  counterclockwise ($\circlearrowleft$, blue dashed) encircling  at different Raman coupling $\Omega_0$. The values $\Omega_0/E_0=0.41,0.82$ and $1.37$ correspond to panels (b),(c) and (d), respectively. Purple dots indicate the initial points, green diamonds the final points, and the gray curves show the branch cuts. (e) The relation between  ${\cal B}$ and the Raman coupling strength $\Omega_0$. The gray dashed lines correspond to the cases shown in (b)(c)(d).
(f) The relation between ${\cal B}$ and the encircling time $T$. The gray dashed lines ($B=1$ and $B=6$) denote the lower dressed $s$ and the upper dressed $d$ bands. For panels (b)-(e), we fix $T E_0/\hbar=7.5$, while for panel (f), $\Omega_0/E_0=0.82$. All other parameters are identical to those in Fig.~\ref{Fig2}, with $\rho/E_0=2.34$.}
\label{Fig3}
\end{center}
\end{figure}

{\it  \color{blue}{Chiral dynamics.---}}
We study chiral dynamics over closed paths in the parameter space via the  the time-dependent Schr\"{o}dinger equation $
i\hbar \partial_t \psi={\cal H}_s(t)\psi$,  where 
 \begin{align}
{\cal H}_s(t)=\sum_{\sigma}K_{\sigma}({\bf r})+\Omega_0 I(r) \sigma_{x}-\omega(t) \sigma_z,
\end{align}
with  $\omega(t)=h(t)-i \gamma(t)$ indicating the closed path in the parameter space.  
Under ${\cal H}_s(t)$, dynamics in different $m$ sectors are decoupled, and their Riemann surfaces possess similar structures [see Supplemental Material]. We therefore focus on the $m=0$ sector, without loss of generality. Therein, the dynamics are governed by $i \hbar \partial_{t} \Phi(t) = {\cal H}(t) \Phi(t)$, where ${\cal H}(t)$ and $\Phi(t)$ denote the non-Hermitian Hamiltonian and the time-evolved state in the $m=0$ sector, respectively [see Supplemental Material for their explicit expressions]. 

To characterize the chiral dynamics,  we expand the time-evolved state as $\Phi (t)=\sum_{\alpha}c_{\alpha}(t) r_{\alpha}(t)$,  where $c_{\alpha}(t)$ are the coefficients and  $r_{\alpha}(t)$ are the instantaneous eigenstates with eigenvalues $E_{\alpha}(t)$.  Chirality is revealed through the spectral trajectory $ {\cal E}(t)$ and the final-state band index ${\cal B}$, defined respectively as 
\begin{align}
 {\cal E}(t)&=\frac{\sum_{\alpha} E_{\alpha}(t) |c_{\alpha}(t)|^2  }{\sum_{\alpha}|c_{\alpha}(t)|^2}, \\
   {\cal B}&=\frac{\sum_{\alpha} \alpha  |c_{\alpha}(T)|^2 }{\sum_{\alpha} |c_{\alpha}(T)|^2},
\end{align}
where $T$ is the time for the system to traverse the closed
path. Intuitively, if the final state predominantly occupies a single dressed band with index $\alpha$, ${\cal B} \approx \alpha$ is close to an integer.

 As illustrated in Fig.~\ref{Fig3}(a), we initialize the system at  the branch cut of the upper $s$ dressed band (with $\alpha=2$), and parameterize the path as
\begin{align}
h(t)= \rho \sin \phi(t), \quad  \gamma(t)=\rho+\rho \cos \phi(t),
\label{loop1}
\end{align} 
 where $\phi(t)=\pm 2\pi t/T+ \pi$ (with $t\in [0,T]$),  and the sign $\pm$ denotes the chirality (clockwise or counterclockwise) of the encircling dynamics. Here  $\rho$ represents the encircling radius. 
 
Figure~\ref{Fig3}(b)(c)(d) show the spectral trajectories, against
the Riemann surface,  with different $\Omega_0$, for the clockwise (red) and counterclockwise (blue) encircling directions, respectively.  In the case with weak Raman couplings  where different dressed bands are well separated, the initial state
evolves primarily within its own band while traversing different Riemann sheets~[see Fig.~\ref{Fig2}(b)]. As is typical for EP encircling~\cite{Rotter-pra-15,Hassan-prl-17}, different encircling directions yield different final states. Here, in a clockwise encircling, the time evolution adiabatically follows the Riemann surface, and the state flips to the lower branch of the dressed $s$ band on returning to the initial parameters. By contrast, in a counterclockwise encircling, a non-adiabatic jump occurs, and the system returns to the initial state at time $T$. The observed chiral dynamics is attributed to the imaginary components of the eigenspectrum, with the time-evolved state tending toward the eigenstate with ${\rm Im}(E_{\alpha}(t))>0$ during the evolution~\cite{Rotter-16,Chan-18, Yang-19,Sun-pra-23}.

This picture is modified when the Raman-induced band hybridization becomes significant.  As shown in Fig.~\ref{Fig3}(c) for $\Omega_0/E_0=0.82$,  while the trajectory remains in the dressed $s$ band under the clockwise encircling,  under a counterclockwise encircling, the trajectory shifts from the upper dressed $s$ to $d$ bands.  More remarkably, further increasing the Raman coupling strength would change the final state in both directions. This is shown in Fig.~\ref{Fig3}(d), where the clockwise (counterclockwise) encircling ends in the lower (upper) dressed  $d$ band.  These observation suggest that the chiral dynamics and its final states are tunable in our setup.

In Fig.~\ref{Fig3}(e), we plot the final-state band index ${\cal B}$ as a function of
the Raman coupling strength for both encircling directions.
For $\Omega_0/E_0 \in (0.25,0.5)$, the chiral dynamics reproduces similar results as in a generic two-level system~\cite{Rotter-pra-15,Hassan-prl-17}. Here the geometry close to the intra-band EP dominates.  For $\Omega_0/E_0 \in (0.75,1.25)$, the chiral dynamics is similar to the
case in Fig.~\ref{Fig3}(c), clockwise encirclings transfer the initial state from the upper to the lower dressed $s$ band,  while counterclockwise encircling ends up in the upper dressed $d$ band. For even strong Raman couplings, clockwise encircling breaks adiabaticity as well, transferring the initial state from the upper dressed $s$ band to the lower dressed $d$ band.  Finally, for sufficiently large $\Omega_0$,  ${\cal B}$ increases significantly,  as final states in both directions occupy  higher-lying dressed bands. This
is due to the fact that the eigenvalues of these higher-lying bands feature larger imaginary components.

We also examine the encircling process with a fixed
path but with increasing encircling time $T$. As shown in Fig.~\ref{Fig3}(f),  for intermediate encircling time  $T E_0 /\hbar \in (5,17.5)$,  
and a counterclockwise rotation, the final state dominantly occupies the upper dressed $d$  band, with ${\cal B}=6$.  By contrast, the final state flips to the lower  dressed $s$ band (with $B= 1$) under a clockwise rotation. Again, in the long-time limit,
the occupation of the higher-lying dressed bands are exponentially occupied due to their larger imaginary eigenvalue components.

\begin{figure}[t]
\begin{center}
\includegraphics[width=0.46\textwidth]{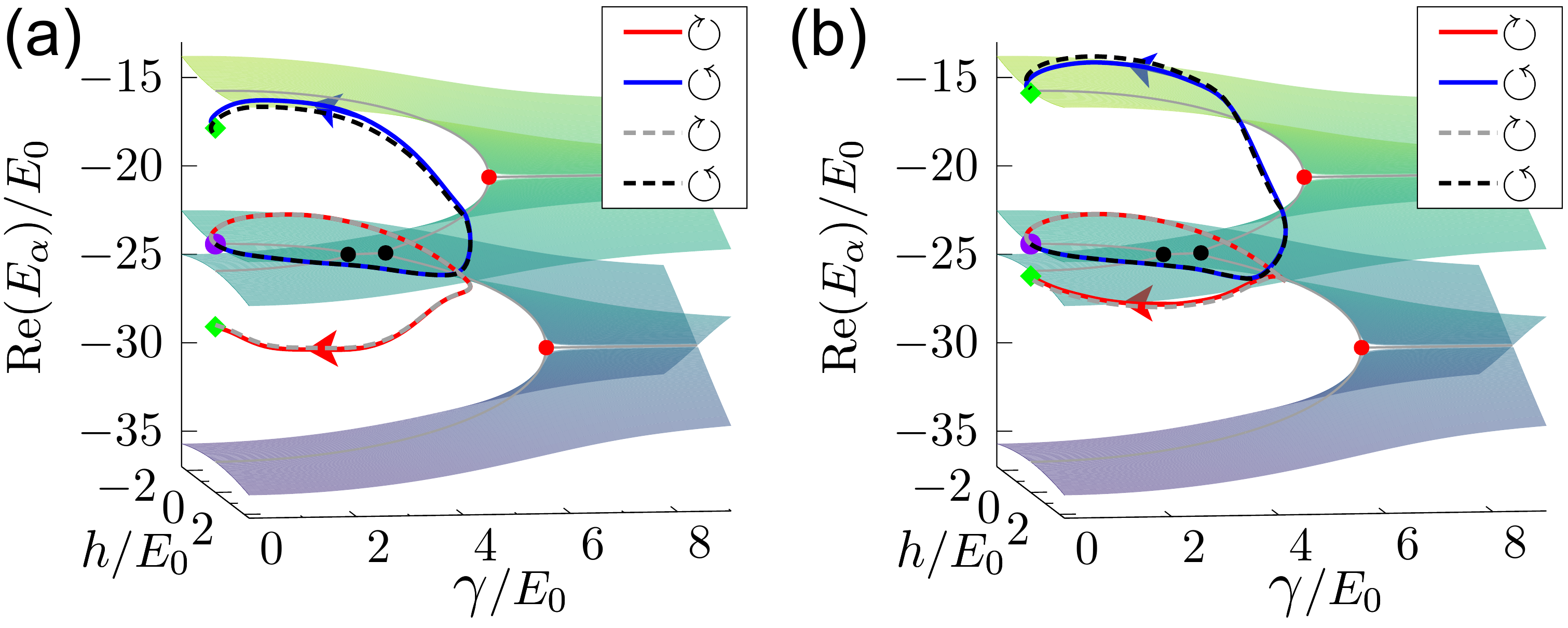}
\caption{Comparison of trajectories calculated from the 
effective and full Hamiltonians, for (a) $T E_0/\hbar=4$ and (b) $T E_0/\hbar=7.5$.  The red (blue) solid lines denote the results of the two-band Hamiltonian for clockwise (counterclockwise) encircling, while the dark–gray (black-gray) dashed lines show the corresponding full-band results.  Here, $\Omega_0/E_0=1.37$ and $\rho/E_0=2.34$. Other parameters are the same as those in Fig.~\ref{Fig2}.}
\label{Fig4}
\end{center}
\end{figure}

{\it  \color{blue}{Effective Model.---}}
As discussed above, except at long times, the overall chiral dynamics involve only two bands, and hence can be captured by an effective two-band model.  Owing to the steep ac Stark potential,  we expand $\psi_{\sigma}({\bf r})=\sum_{m,n=s,d} \varphi_{mn \sigma}(r) \Theta_m(\theta) a_{mn \sigma}$,  restricting the basis to the $s$ and $d$ bands. This choice is motivated by two considerations:  the initial state occupies the upper dressed $s$ band; and the coupling between the $s$ and $d$ bands is much stronger than that between the $s$ and $p$ bands.  Within these factors,  we derive  an effective Hamiltonian in the $m=0$ sector,  where $H_{\rm{eff}}= \Phi^{\dag}  {\cal H}_{\rm{eff}}\Phi$ and 
\begin{align}
{\cal H}_{{\rm eff}}=\left(\begin{array}{cccc} \epsilon_{s \uparrow}-\omega & \Omega_{s\uparrow,s\downarrow} & 0 & \Omega_{s\uparrow,d\downarrow} \\ \Omega_{s\uparrow,s\downarrow} & \epsilon_{s \downarrow}+\omega & \Omega_{d\uparrow,s\downarrow} & 0 \\0 & \Omega_{d\uparrow,s\downarrow} & \epsilon_{d \uparrow}-\omega & \Omega_{d\uparrow,d\downarrow} \\ \Omega_{s\uparrow,d\downarrow} & 0 & \Omega_{d\uparrow,d\downarrow} & \epsilon_{d \downarrow}+\omega
\end{array}\right).
\label{Heff}
\end{align}
Here $\Phi=\left(\begin{array}{cccc}a_{s \uparrow} & a_{s\downarrow}& a_{d \uparrow}& a_{d\downarrow}\end{array}\right)^{T}$.  This effective Hamiltonian captures both the static and dynamical properties of the system. To examine the validity of the effective Hamiltonian,  we compare the trajectories calculated from dynamics under the effective and full Hamiltonians. As
shown in Fig.~\ref{Fig4}, the results agree remarkably well at different intermediate time scales. 

Under the effective model,  the impact of the encircling time $T$, particularly in the non-adiabatic evolution in Fig.~\ref{Fig3}(d),  can be well explained. For example,  under an intermediate encircling time $T$ [see Fig.~\ref{Fig4}(b)], the final state in either direction is transferred to the dressed $d$ band. This behavior arises from the different imaginary eigenvalue components of the bands. These can be obtained by analyzing Eq.~(\ref{Heff}).  When the coupling between the $s$ and $d$ bands vanishes,  Eq.~(\ref{Heff}) reduces to 
${\cal H}_{\rm{eff}}={\cal M}_{s}\bigoplus {\cal M}_{d}$, where 
\begin{align}
 {\cal M}_n=\left(\begin{array}{cc} \epsilon_{n \uparrow}-\omega & \Omega_{n\uparrow, n \downarrow} \\ \Omega_{n\uparrow,n \downarrow} & \epsilon_{n \downarrow}+\omega \end{array}\right).
\label{Heff_single}
\end{align}
The eigenvalues of Eq.~(\ref{Heff_single}) are $E_{\pm}=\epsilon_{n}\pm \sqrt{\omega^2+(\Omega_{n \uparrow, n \downarrow})^2}$, with  $\epsilon_{n \uparrow}=\epsilon_{n \downarrow}=\epsilon_n$ for $m=0$. Since $|\Omega_{s\uparrow,s \downarrow}|> |\Omega_{d \uparrow, d \downarrow}|$,  the imaginary eigenvalue components
of the $d$ band is larger than that of the $s$ band based on Eq.~(\ref{Heff_single}).  Hence, when increasing the encircling time $T$,  there exists a time window in which the occupation of the $d$ band becomes dominant,  as observed in Fig.~\ref{Fig3}(d).

\begin{figure}[t]
\begin{center}
\includegraphics[width=0.46\textwidth]{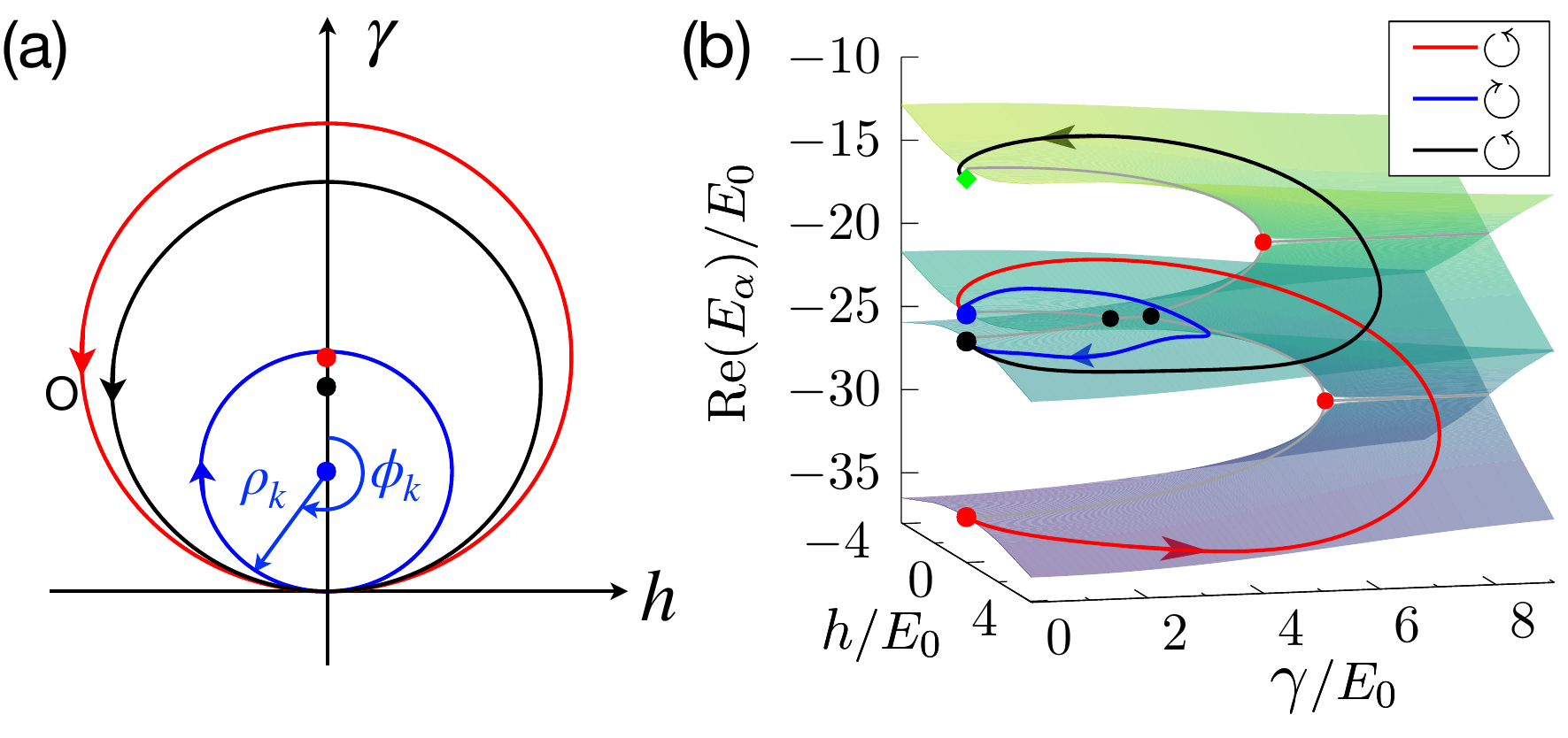}
\caption{(a) Schematic of the designed loops. The red, blue and black circles denotes the first, second and third loops with distinct encircling radius, and the corresponding dots mark their origins. (b) Spectral trajectories of the encircling. Here, $\rho_{k=1,2,3}=(4, 2.05, 3.5)E_0$, $T_{k=1,2,3}=(1.35, 3.2, 1.1)\hbar/E_0$ with $\Omega_0/E_0=1.37$.  Other parameters are the same as those in Fig.~\ref{Fig2}.}
\label{Fig5}
\end{center}
\end{figure}

{\it  \color{blue}{Encircling multiple EPs.---}}
The tunable chiral preparation of states in high-lying bands demonstrated thus far hinges upon the presence
of intra-band EPs, and often involve non-adiabatic processes. But with the emergence of inter-band EPs, one can design an adiabatic pathway toward higher-lying bands on the Riemann surface, making use of both types of EPs.

For this purpose, we consider the path in Fig.~\ref{Fig5}(a),  which consists of several loops parameterized by
\begin{align}
h(t)= \rho_k \sin \phi_k(t), \quad  \gamma(t)=\rho_k+\rho_k \cos \phi_k(t),
\label{loop-adiabatic}
\end{align} 
where  
$\phi_{k}(t)=(-1)^{k} 2\pi \left(t-\sum^{k-1}_{k=1}T_{k-1}\right)/T_k+\pi$  with $k=1,2,3$ and 
 $T_0=0$. 
 Physically, we start from the lower dressed $s$ band,  and the first loop is chosen to encircle the intra-band EP in the dressed $s$ band [see Fig.~\ref{Fig5}(b)]; the second loop encircles the inter-band EP between the dressed $s$ and $d$ bands; and the third loop encircles the intra-band EP of the dressed $d$ band. Figure~\ref{Fig5}(b) shows the resulting spectral trajectory based on the effective Hamiltonian [see Supplemental Material for the comparison between results from the effective and full Hamiltonians]. As expected, an adiabatic evolution along the path, including both clockwise
and counterclockwise sectors, leaves the system in the upper branch of a higher band. Note that while we illustrate the adiabatic chiral state transfer 
using the $s$ and $d$ bands, the design can be extended to high-lying bands (for instance between $p$ and $f$ bands).

{\it \color{blue}{Discussion.---}} 
We have illustrated the parametric chiral state transfer of atoms under the dissipative SOAMC, designed according to the rich exceptional structure of the system. 
The chiral dynamics is useful for the path-dependent preparation of atoms in the high-lying bands. 
While our results are readily accessible in a non-interacting
condensate subject to a dissipative SOAMC, 
we expect the consideration of nonlinear contributions of condensate interactions would further enrich the exceptional structure and chiral dynamics.

{\it  \color{blue}{Acknowlegement.--}} We acknowledge fruitful discussions with Konghao Sun. This work is supported by  the Natural Science Foundation of China (Grants No. 12104406, 12374479, 12204105).
K.C. acknowledges support by Zhejiang Provincial Natural Science Foundation (Grant No. ZCLMS25A0401) and the startup grant of Zhejiang Sci-Tech University (Grant No. 21062338-Y).  
W.Y. acknowledges support from the Innovation Program for Quantum Science and Technology (Grant No. 2021ZD0301904). 

\bibliographystyle{apsrev4-2}
%\normalem

\clearpage

%\newpage

\widetext
\begin{center}
\textbf{\large Supplemental Material for  ``Chiral Dynamics Near Intra- and Inter-Band Exceptional Points under Dissipative
Spin-Orbital-Angular-Momentum Coupling"}
\end{center}
%%%%%%%%%% Merge with supplemental materials %%%%%%%%%%
%%%%%%%%%% Prefix a "S" to all equations, figures, tables and reset the counter %%%%%%%%%%
\setcounter{equation}{0}
\setcounter{figure}{0}
\setcounter{table}{0}
\makeatletter
\renewcommand{\theequation}{S\arabic{equation}}
\renewcommand{\thefigure}{S\arabic{figure}}
\renewcommand{\citenumfont}[1]{#1}
%%%%%%%%%% Prefix a "S" to all equations, figures, tables and reset the counter %%%%%%%%%%

In this Supplemental Material, we present details of the eigenspectrum of a non-Hermitian Hamiltonian obtained from both the direct basis expansion and the full-band expansion. We also provide the chiral dynamical formalism and compare the adiabatic encircling results from the effective two-band model with those from full-band calculations.

\section{eigenspectrum of a non-Hermitian Hamiltonian}
\subsection{Direct basis expansion}
As discussed in the main text,  the eigenspectrum of ${\cal H}_s$,  as shown in Eq.~(\ref{Hs}),  can be obtained directly by choosing an appropriate basis.     
Specifically,  the eigenvalues are determined by diagonalizing the following Schr\"odinger equation,  
\begin{eqnarray}
\begin{split}
\left(\begin{array}{cc}K_{\uparrow }({\bf r})-\omega & \Omega (r) \\\Omega (r) & K_{\downarrow }({\bf r})+\omega \end{array}\right) 
\left(\begin{array}{c}u_\eta({\bf r}) \\v_{\eta}({\bf r})\end{array}\right)
 =E_\eta \left(\begin{array}{c}u_{\eta}({\bf r}) \\v_{\eta}({\bf r})\end{array}\right), 
\label{supp-matrix}
\end{split}
\end{eqnarray}
where $E_\eta \in \mathbb C$ are the eigenvalues with $\eta$ labeling the energy index, and $u_{\eta}({\bf r})$ and $v_{\eta}({\bf r})$ are the corresponding eigenstate components.  Here, $\omega=h-i \gamma$, where $h$  denotes the two-photon detuning and $\gamma$ characterizes the atom loss. In Eq.~(\ref{supp-matrix}),  $K_{\sigma}({\bf r})=-\frac{\hbar^2}{2M}\Big[\frac{1}{r}\frac{\partial}{\partial r}\left(r\frac{\partial}{\partial r}\right)+\frac{1}{r^2}\left(\frac{\partial}{\partial \theta}-i l \tau \right)^2\Big]+\chi (r)$ with $\tau=+1(-1) $ for  $\sigma=\uparrow(\downarrow)$.  As shown in the main text,   the Raman coupling $\Omega(r)$ and the ac Stark shift $\chi (r)$ take the forms $\Omega_0 I(r)$ and $\chi_0 I(r)$,  where $\Omega_0$ and $\chi_0$ denote their corresponding strengths, respectively.   The intensity profile is given by $I(r) =   (\sqrt{2}r/w)^{|l_1|+|l_2|}e^{-2 r^2 /w^2}$ with $w$ the beam waist.  Here,  we impose an isotropic hard-wall boundary of radius $R$ in realistic numerical calculations. 

\begin{figure}[t]
\begin{center}
\includegraphics[width=0.8\textwidth]{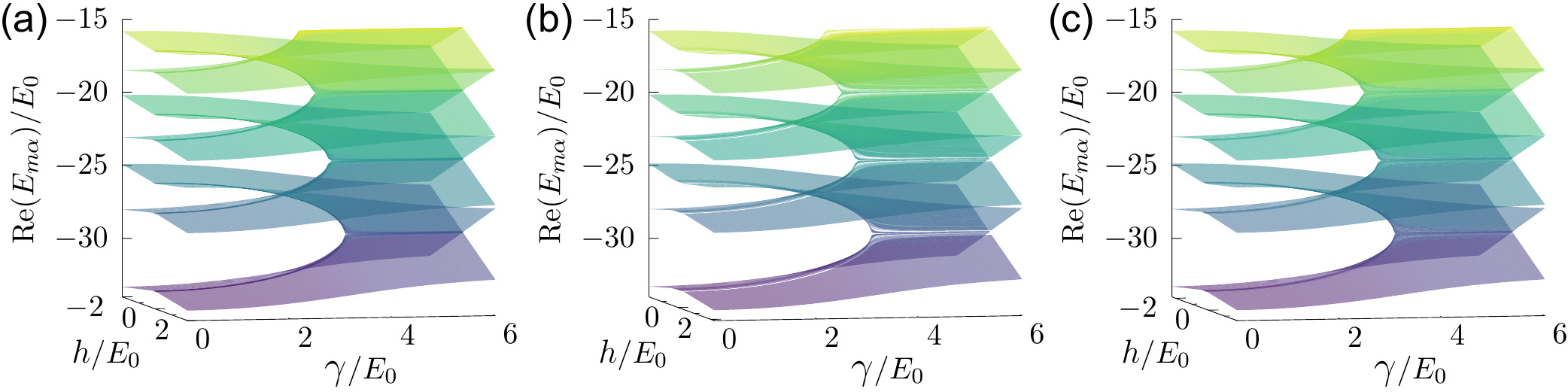}
\caption{Riemann structure of real parts of the eigenvalues in the $h-\gamma $ plane for different $m$. Here,  (a) $m=0$, (b) $m=-1$ and (c) $m=1$ with  fixed $\Omega_0/E_0=0.82$. Other parameters are the same as those in Fig.~\ref{Fig2}.}
\label{FigS1}
\end{center}
\end{figure}

Considering  $[L_z, {\cal H}_s]=0$,  with $L_z=-i \hbar \partial /\partial \theta$,  and imposing an isotropic two-dimensional hard-wall boundary condition, the radial wave function $f_{m \alpha \sigma}(r)$ can be expanded in terms of the Bessel functions, where $\alpha$ labels the radial eigenmode.   Accordingly,  the eigenstate of ${\cal H}_s$ can be written as 
 
\begin{align}
u_\eta({\bf r}) &=f_{m\alpha \uparrow}(r)\Theta_{m}(\theta)= \sum_{p}c^{p}_{m \alpha\uparrow}R_{p,m-l}(r)\Theta_{m}(\theta), \label{supp-un}\\
v_\eta({\bf r}) &=f_{m\alpha  \downarrow}(r)\Theta_{m}(\theta)=\sum_{p}c^{p}_{ m\alpha \downarrow}R_{p,m+l}(r)\Theta_{m}(\theta),  \label{supp-vn}
\end{align}
where  $R_{pm}(r)=\sqrt{2}J_{m}(z_{pm} r/R)/\left[R J_{m+1}(z_{pm}) \right]$ and $\Theta_{m}(\theta)=e^{im\theta}/\sqrt{2\pi}$. Here $R_{pm}(r) $ is the radial wave function of the two-dimensional hard-wall potential,  and $z_{pm}$ is the $p$th zero of $J_{m}(x)$ with $p\in \mathbb{Z}^{+}$.   Substitute  Eq.~(\ref{supp-un}) and (\ref{supp-vn})  into Eq.~(\ref{supp-matrix}),  each $m$ becomes decoupled and in each $m$ sector, we have
\begin{eqnarray}
\sum_{p'} \left(\begin{array}{cc}K^{pp'}_{\uparrow,m-l} & \Omega^{p', m+l}_{p,m-l} \\\Omega^{p',m-l}_{p,m+l} & K^{pp'}_{\downarrow,m+l}\end{array}\right)
\left(\begin{array}{c}c^{p'}_{m\alpha \uparrow} \\c^{p'}_{m\alpha \downarrow}\end{array}\right)& = & E_{m\alpha} \left(\begin{array}{c}c^{p}_{ m\alpha \uparrow} \\c^{p}_{m\alpha \downarrow}\end{array}\right),
\label{Hamiltonian}
\end{eqnarray}
where $K^{pp'}_{\sigma,m-l\tau}$  and $\Omega^{p',m'}_{p,m}$  have the following forms, 
\begin{eqnarray}
K^{pp'}_{\sigma,m-l\tau}  &=&  \left(\frac{\hbar^2 z^2_{p,m-l\tau}}{2MR^2}-\omega \tau \right)\delta_{pp'}+\chi_0 \int r dr R_{p,m-l\tau} I(r) R_{p',m-l\tau}, \label{Kmatrix}\\
\Omega^{p',m'}_{p, m}&=&\Omega_0 \int rdr R_{p,m}(r) I(r) R_{p',m'}(r).
\label{Omega1}
\end{eqnarray}
Diagonalizing Eq.~(\ref{Hamiltonian}) yields the eigenvalues and eigenfunctions. Figure~\ref{FigS1} shows the real parts of the eigenspectra for different $m$ at a fixed Raman coupling. A common feature is the emergence of exceptional points (EPs), and the overall Riemann-surface structures remain similar across different $m$, despite variations in the EP values.
 
While the above approach is straightforward for obtaining the eigenspectrum of ${\cal H}_s$, its physical meaning can become ambiguous, particularly when extended to dynamics. To clarify the physics, we adopt an alternative approach.

\subsection{Full-band expansion}
To capture the underlying physics,  as discussed in the main text, the eigenspectrum of Hamiltonian (\ref{Hs}) can  be derived more transparently by reconsidering the non-Hermitian Hamiltonian,
\begin{eqnarray}
H_0 & = &  \int d {\bf r}
\left(\begin{array}{cc}\psi^{\dag}_{\uparrow}({\bf r}) & \psi^{\dag}_{\downarrow}({\bf r})\end{array}\right)
\left(\begin{array}{cc}K_{\uparrow}({\bf r})-\omega & \Omega_0 I(r) \\ \Omega_0 I(r) & K_{\downarrow}({\bf r})+\omega \end{array}\right)
\left(\begin{array}{c}\psi_{\uparrow}({\bf r}) \\ \psi_{\downarrow}({\bf r})\end{array}\right).
\label{supp-H0}
\end{eqnarray}
Inspired by Ref.~\cite{supp-Wang-21, supp-Chen-prr-22},  we expand the field operator  $\psi_\sigma(\mathbf{r})$ in two dimensions  as
\begin{align}
   \psi_{\sigma}({\bf r})=\sum_{m,n} \varphi_{mn \sigma}(r) \Theta_m(\theta) a_{mn \sigma},
 \label{band-expansion}
\end{align}
where $\varphi_{m n \sigma}(r)$ satisfies 
\begin{align}
\label{Ksigma}
 K_{\sigma}({\bf r}) \varphi_{m n \sigma}(r)\Theta_m(\theta)=\epsilon_{mn\sigma}\varphi_{mn\sigma}(r)\Theta_m(\theta),
\end{align}
as shown in the main text. Here,  $n$ and $m$ denote the radial and angular quantum numbers, respectively. Specifically,   the radial wave function $\varphi_{mn\sigma}(r)$ can be expanded by the radial function of the two-dimensional hard-wall potential,
\begin{align}
\varphi_{mn\sigma}(r)=\sum_{p}c^{p}_{mn \sigma} R_{p,m-l \tau}(r), 
\label{supp-cmn}
\end{align}
where $\tau=+1(-1)$ for $\sigma=\uparrow (\downarrow)$.  Substitute Eq.~(\ref{supp-cmn}) to Eq.~(\ref{Ksigma}),  we have the following equation
\begin{eqnarray}
\sum_{p'} K^{pp'}_{\sigma,m-l\tau} c^{p'}_{ mn \sigma}  & = & \epsilon_{mn \sigma}  c^{p}_{ mn \sigma},
\label{supp-Kmatrix}
\end{eqnarray}
where $K^{nn'}_{\sigma,m-l\tau}$ can be found in Eq.~(\ref{Kmatrix}).  Solving Eq.~(\ref{supp-Kmatrix}),  $\epsilon_{mn\sigma}$ and  the ridial wave function of $K_{\sigma}({\bf r})$ can be obtained.  Intuitively, when the ac Stark potential becomes steep, i.e., $\chi_0 \ll 0$,  a large energy gap emerges between neighboring $n$ levels for a given angular quantum number$m$.  In this scenario,  the index  $n$  plays a role analogous to the band index, i.e., $n=s,p,d,f, \cdots$.  

\begin{figure}[t]
\begin{center}
\includegraphics[width=0.8\textwidth]{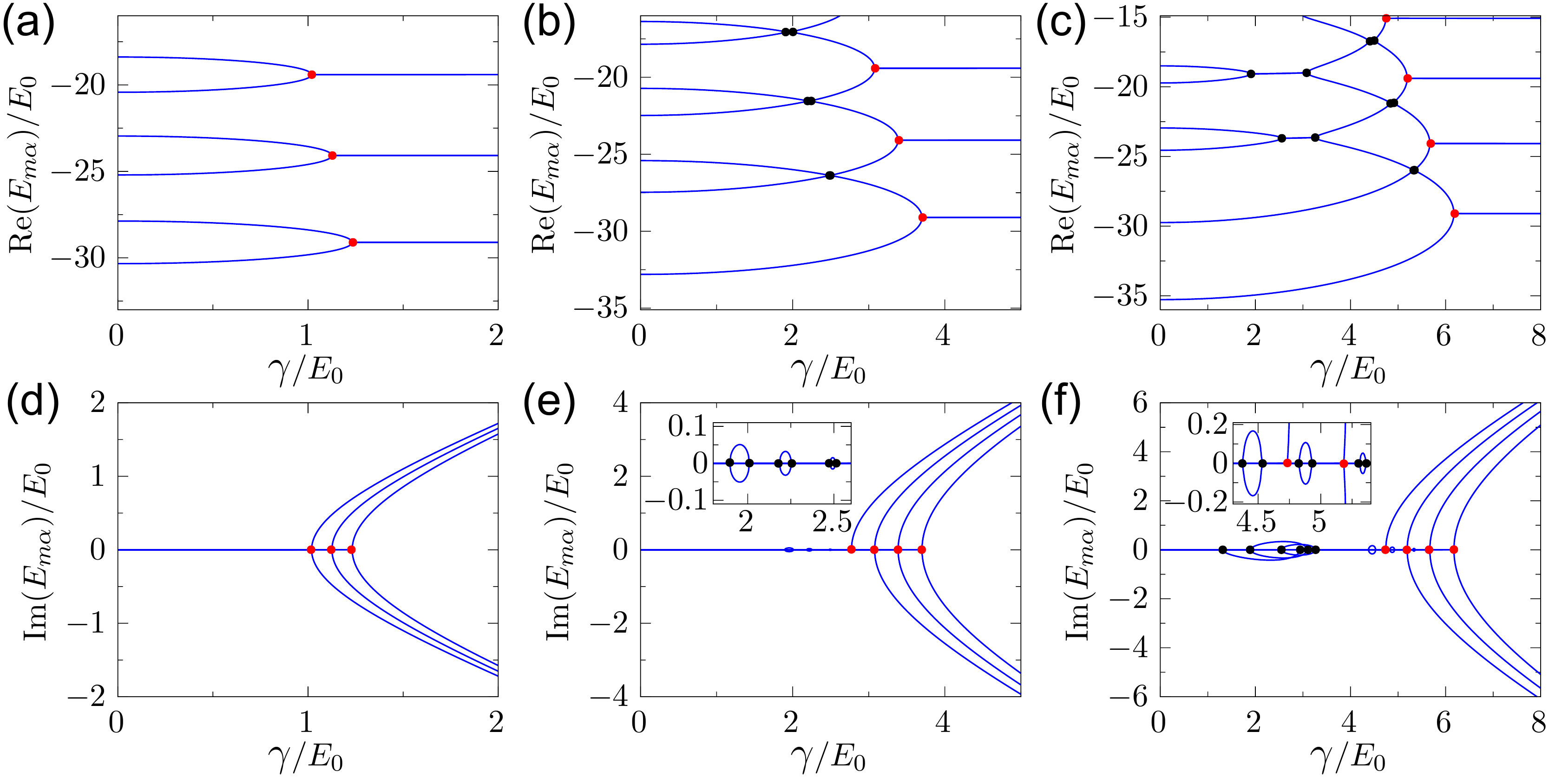}
\caption{Real and imaginary parts of the eigenspetra as $\gamma$ increases for $h=0$ under different Raman coupling $\Omega_0$ in the $m=0$ sector. Here, (a)(b)(c) denote the real part of the eigenspetra for $\Omega_0/E_0=0.27,0.82$ and $1.37$. (d)(e)(f) are the corresponding imaginary parts, the inset plots in panels (e) and (f) provide zoomed-in views. Other parameters are the same as those in Fig.~\ref{Fig2}.}
\label{FigS2}
\end{center}
\end{figure}

Substitute Eq.~(\ref{band-expansion}) into Eq.~(\ref{supp-H0}),  we have $H_0=\sum_{m}H_{m}$, where $H_m$ is given by 
\begin{align}
\label{Hamiltonian-eff}
H_{m}=\sum_{n\sigma } \Big[\epsilon_{mn\sigma}-\omega\tau \Big]a^{\dag}_{mn \sigma}a_{mn\sigma}+\sum_{nn'}\Big(\Omega^{m}_{n\uparrow, n' \downarrow} a^{\dag}_{mn \uparrow}a_{m n' \downarrow}+h.c.\Big),
\end{align}
Here, we define 
\begin{eqnarray}
\Omega^{m}_{n \sigma, n'\sigma'} & = & \Omega_0 \int rdr \varphi_{m n \sigma} I(r) \varphi_{mn' \sigma' }.
\label{Omega-m}
\end{eqnarray}
Diagonalizing Eq.~(\ref{Hamiltonian-eff}) yields and the eigenvalues and eigenstates. Figure~\ref{FigS2} shows the real and imaginary parts of the eigenspectra in the $m=0$ sector for $h=0$ as the Raman coupling $\Omega_0$ increases. As discussed in the main text, for weak Raman coupling, the real dressed bands are well separated, and each hosts an intra-band EP (marked in red) [see Fig.~\ref{FigS2}(a)]. As the Raman coupling strength increases, different dressed bands hybridize, leading to the emergence of inter-band EPs [see Fig.~\ref{FigS2}(b)]. Correspondingly,  the number of EPs increases with the coupling strength [see Fig.~\ref{FigS2}(c)].  Figure~\ref{FigS2} also shows the corresponding imaginary dressed bands as the Raman coupling $\Omega_0$ increases, which are consistent with the real-band structures.

Compared with the direct basis expansion, the full-band expansion provides clear physical insight into both the static and dynamical properties; in particular, it enables the derivation of an effective Hamiltonian that elucidates the underlying physics. For example,  as discussed in the main text,  restricting the expansion to  the $s$ and $d$ bands reduces  Eq.~(\ref{Hamiltonian-eff}) to 
$H_{m}=\Phi^{\dag}_m  {\cal H}_m\Phi_m$, where 
\begin{align}
{\cal H}_m=\left(\begin{array}{cccc} \epsilon_{ms \uparrow}-\omega & \Omega^m_{s\uparrow,s\downarrow} & 0 & \Omega^m_{s\uparrow,d\downarrow} \\ \Omega^m_{s\uparrow,s\downarrow} & \epsilon_{ms \downarrow}+\omega & \Omega^m_{d\uparrow,s\downarrow} & 0 \\0 & \Omega^m_{d\uparrow,s\downarrow} & \epsilon_{md \uparrow}-\omega & \Omega^m_{d\uparrow,d\downarrow} \\ \Omega^m_{s\uparrow,d\downarrow} & 0 & \Omega^m_{d\uparrow,d\downarrow} & \epsilon_{md \downarrow}+\omega
\end{array}\right),
\label{supp-effective}
\end{align}
and $\Phi_m=\left(\begin{array}{cccc}a_{m s \uparrow} & a_{ms\downarrow}& a_{md \uparrow}& a_{md\downarrow}\end{array}\right)^{T}$. Retaining only the $n$th term yields the one-band approximation, in which case, Eq.~(\ref{Hamiltonian-eff}) reduces to  $H_m=\Phi^{\dag}_m {\cal H}_m \Phi_m$, where 
\begin{align}
{\cal H}_m=
\left(\begin{array}{cc} \epsilon_{mn \uparrow}-\omega & \Omega^m_{n\uparrow,n\downarrow} \\ \Omega^m_{n \uparrow,n \downarrow} & \epsilon_{m n \downarrow}+\omega
\end{array}\right),
\label{supp-Hm}
\end{align}
and $\Phi_m=\left(\begin{array}{cc}a_{m n \uparrow} & a_{mn\downarrow}\end{array}\right)^{T}$. When we focus on the $m=0$ sector,  Eq.~(\ref{supp-effective}) and (\ref{supp-Hm}) reduce to Eq.~(\ref{Heff}) and (\ref{Heff_single}) in the main text. These effective Hamiltonians capture both the static and dynamical properties of the system.
  
\section{chiral dynamical formalism} 
The chiral dynamics are studied via the time-dependent Schr\"{o}dinger equation,  $ i\hbar \partial_t \psi={\cal H}_s(t)\psi$,  where 
 \begin{align}
{\cal H}_s(t)=\sum_{\sigma}K_{\sigma}({\bf r})+\Omega_0 I(r) \sigma_{x}-\omega(t) \sigma_z,
\end{align}
with  $\omega(t)=h(t)-i \gamma(t)$.  Since $[L_z,{\cal H}_s(t)]=0$,  we expand $\psi_{\sigma}({\bf r},t)=\sum_{m,n} \varphi_{mn \sigma}(r) \Theta_m(\theta) a_{mn \sigma}(t) $, which reduces the dynamics  to 
$ i \hbar \partial_{t} \Phi_{m}(t) = {\cal H}_m (t) \Phi_{m}(t)$, as  ${\cal H}_s(t)$ decouples into independent $m$ sectors. Here ${\cal H}_{m}(t)$ is the instantaneous non-Hermitian Hamiltonian in the $m$ sector and $\Phi_{m}(t)$ is the corresponding time-evolved state, with explicit expressions for both given by
\begin{align}
\label{Hm_matrix}
    {\cal H}_m(t)=
    \left(\begin{array}{ccccc} \epsilon_{ms \uparrow}-\omega(t)  & \Omega^{m}_{s\uparrow,s\downarrow} & 0 & \Omega^{m}_{s \uparrow, p\downarrow } & \cdots  \\ \Omega^m_{s \uparrow, s \downarrow} & \epsilon_{ms \downarrow}+\omega(t) & \Omega^m_{p\uparrow,s \downarrow} & 0 & \cdots \\  0 & \Omega^{m}_{p\uparrow,s\downarrow} & \epsilon_{mp \uparrow}-\omega(t) & \Omega^m_{p\uparrow, p\downarrow} & \cdots \\\Omega^m_{s\uparrow,p\downarrow} & 0 & \Omega^m_{p\uparrow,p\downarrow} & \epsilon_{ms \downarrow}+\omega(t) & \cdots \\ \cdots  & \cdots & \cdots & \cdots & \ddots \end{array}\right),
\end{align}
and
$\Phi_m(t)= \left(\begin{array}{ccccc} a_{ms\uparrow}(t) & a_{ms \downarrow}(t) & a_{mp\uparrow}(t) & a_{mp \downarrow}(t) & \cdots \end{array}\right)^{T}$,  where the defination of $\epsilon_{mn \sigma}$  and $\Omega^{m}_{n\sigma,n'\sigma'}$ can be found in  Eq.~(\ref{Ksigma}) and (\ref{Omega-m}). Thus, given an intial state $\Phi_m(0)$ and the time evolution of $\omega(t)$,  the time-evolved state  $\Phi_m(t)$ can be obtained.

As discussed in the main text, we focus on the $m=0$ sector. In this case, ${\cal H}_m(t)$ and $\Phi_m(t)$ reduces to
\begin{align}
    {\cal H}(t)=
    \left(\begin{array}{ccccc} \epsilon_{s \uparrow}-\omega(t)  & \Omega_{s\uparrow,s\downarrow} & 0 & \Omega_{s \uparrow, p\downarrow } & \cdots  \\ \Omega_{s \uparrow, s \downarrow} & \epsilon_{s \downarrow}+\omega(t) & \Omega_{p\uparrow,s \downarrow} & 0 & \cdots \\  0 & \Omega_{p\uparrow,s\downarrow} & \epsilon_{p \uparrow}-\omega(t) & \Omega_{p\uparrow, p\downarrow} & \cdots \\\Omega_{s\uparrow,p\downarrow} & 0 & \Omega_{p\uparrow,p\downarrow} & \epsilon_{s \downarrow}+\omega(t) & \cdots \\ \cdots  & \cdots & \cdots & \cdots & \ddots \end{array}\right),
\end{align}
and $\Phi(t)= \left(\begin{array}{ccccc} a_{s\uparrow}(t) & a_{s \downarrow}(t) & a_{p\uparrow}(t) & a_{p \downarrow}(t) & \cdots \end{array}\right)^{T}$. Here, we have omitted the angular quantum number for simplicity.

\begin{figure}[t]
\begin{center}
\includegraphics[width=0.5\textwidth]{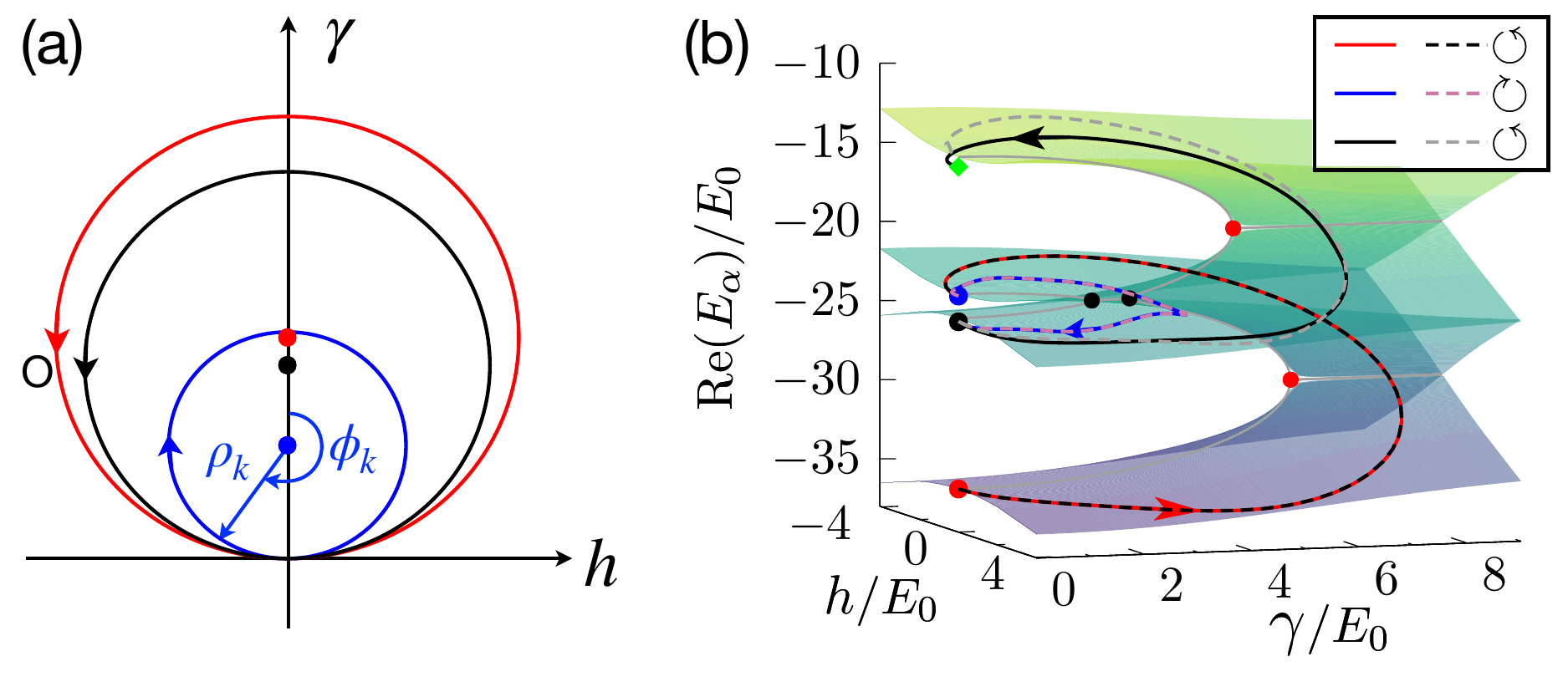}
\caption{Comparison of adiabatic encircling results from the two-band and full-band models. (a) Schematic of the designed loops. (b) Spectral trajectories of the encircling.  The solid lines denote the results from the effective two-band model,  while the dashed lines correspond to the full-band calculation.  Here, $\Omega_0/E_0=1.37$. Other parameters are the same as those in Fig.~\ref{Fig2} and Fig.~\ref{Fig5}.}
\label{FigS3}
\end{center}
\end{figure}

\section{Comparisons of adiabatic encircling}
For the adiabatic encircling discussed in the main text,  we compare the spetral trajectories obtained from the effective two-band model with those from the full-band calculation.  As illustrated in Fig.~\ref{FigS3}, the two results show excellent agreement,  particularly for short encircling times $t$. For large $t$, deviations emerge due to the increasing importance of higher-band contributions, which lie beyond the two-band description.  

\bibliographystyle{apsrev4-2}
%\normalem

\end{document}